\documentclass[reqno]{amsart} 

\usepackage{graphicx}
\usepackage{bm}

 \numberwithin{equation}{section}
\def\stackreb#1#2{\ \mathrel{\mathop{#1}\limits_{#2}}}
\newcommand{\nc}{\newcommand}
\newcommand{\Z}{\mathbb Z}
\nc{\rnc}{\renewcommand} \nc{\beq}{\begin{equation}}
\nc{\eeq}{\end{equation}} \nc{\beqa}{\begin{eqnarray}}
\nc{\eeqa}{\end{eqnarray}}
\def \z{\underline{z}}

\def \y{\underline{y}}
\def \N{\mathcal{N}}

\def\stackreb#1#2{\ \mathrel{\mathop{#1}\limits_{#2}}}
\def \T{\mathbb{T}}

\begin{document}

\title[Chiral symmetry breaking]
{\bf Vanishing superconformal indices and\\
the chiral symmetry breaking}

\author{V.~P.~Spiridonov}
\address{Bogoliubov Laboratory of Theoretical Physics,
JINR, J. Curie str. 6, Dubna, Moscow Region 141980, Russia;
e-mail address: spiridon@theor.jinr.ru \hfill{}
{\em Current address:} Max Planck Institut f\"ur Mathematik,
Vivatsgasse 7, 53111 Bonn, Germany
}

\author{G.~S.~Vartanov}
\address{DESY Theory, Notkestrasse 85, 22603 Hamburg, Germany;
e-mail address: vartanovg@yahoo.com
}

\begin{abstract}
Superconformal indices  of $4d$ $\mathcal{N }=1$ SYM
theories with $SU({N})$ and $SP(2{N})$ gauge groups
are investigated for $N_f={N}$ and $N_f={N}+1$ flavors,
respectively. These indices vanish for generic values of
the flavor fugacities.
However, for a singular submanifold of fugacities they
behave like the Dirac delta functions and describe the
chiral symmetry breaking phenomenon. Similar picture holds
for partition functions of $3d$ supersymmetric
field theories with the chiral symmetry breaking.
\end{abstract}

\maketitle

\tableofcontents

\section{Introduction}

We take as a starting point the remarkable observation of \cite{Dolan:2008qi}
that superconformal indices (SCIs) of $4d$ supersymmetric field theories are
expressed in terms of elliptic hypergeometric integrals (EHIs) discovered
in \cite{S1,S2} (for a review see \cite{S3}). SCIs were introduced  in
\cite{Kinney} and \cite{Romelsberger1,Romelsberger2} from different physical motivations.
They describe also indices of nonconformal supersymmetric field theories on
curved backgrounds flowing to a superconformal infrared fixed point
\cite{FS}. In \cite{Kinney} the main target was the AdS/CFT correspondence.
In \cite{Romelsberger1,Romelsberger2} BPS operators of $\mathcal{N }=1$ SYM theories were
studied and the equality of SCIs for Seiberg dual theories was conjectured.
In \cite{Dolan:2008qi} this hypothesis was proven analytically for the
initial Seiberg duality \cite{Seiberg:d} using mathematical properties
of EHIs established in \cite{S1,S2,Rains,spi:short}.
Following this result we systematically considered the connection of
$\mathcal{N }=1$ supersymmetric field theories with the theory of
EHIs \cite{SV1}-\cite{SV5}
We showed that available physical
checks for Seiberg dualities can be described by known general
properties of EHIs, conjectured many new mathematical identities
and found many new physical dualities. SCI techniques was applied also to the
description of $S$-dualities of $\mathcal{N }=2,4$ extended supersymmetric
field theories in \cite{GPRR1}-\cite{GRRY2} and \cite{SV4}.

In \cite{SV3} it was conjectured that all 't Hooft anomaly
matching conditions are related to the total ellipticity condition
for EHIs \cite{S3}. As shown in \cite{sudano} this is not so for
$U(1)_R$ and $U(1)_R^3$-anomalies. However, in \cite{SV5} it was
demonstrated that all anomaly matchings for Seiberg dual theories
follow from $SL(3,\mathbb{Z})$-modular
transformation properties of the kernels of dual SCIs. One can
consider modifications of SCIs such as the addition
of charge conjugation \cite{Zwiebel:2011wa}, inclusion of surface
operators \cite{Nakayama:2011pa,GRR}
or line operators \cite{Dimofte:2011py}-\cite{Gang:2012yr2}, etc.
Connection of SCIs of $4d$ theories and partition functions of $2d$
statistical mechanics models was discussed in \cite{S5}. An interesting
$5d/4d$ boundary field theory with the extended $E_7$-flavor symmetry
was proposed in \cite{DG}, which is based on the particular $4d$
multiple dual theories \cite{SV1} and $W(E_7)$-symmetry of corresponding
SCIs \cite{S2,Rains}. A similar interpretation for
SCIs with $W(E_6)$-symmetry \cite{SV1} was proposed in \cite{GV}.

In this paper we would like to discuss a particular phenomenon for
$\N=1$ SYM theories pointed out in \cite{Seiberg} which is
known as the confinement with chiral symmetry breaking when the global
symmetry group gets broken. Originally such a physical effect was
considered for $SU({N})$-gauge group supersymmetric quantum chromodynamics
with $N_f={N}$ and it
should be contrasted to the so-called $s$-confinement occurring
at $N_f={N}+1$ (i.e., the confinement without breaking global symmetries).
Later the theories with quantum modified moduli space were systematically
studied in \cite{Grinstein:1997zv,Grinstein:1997zv2}. SCIs of such theories  are not well
defined, in particular, they were not computed from the first principles
using the localization techniques. The formal free field computations
yield diverging results (actually, such computations are not affected
even by the anomalies \cite{SV3b,SV5}).
From mathematical point of view the relevant behavior of SCIs was  partially
considered for ${N}=2$ in \cite{spi:cont}. We compute SCIs of such theories
in general case $N_f={N}$ and show that they involve Dirac delta
functions reflecting the presence of chiral symmetry breaking.
Similar considerations are fulfilled for a $4d$ theory with
$SP(2{N})$ gauge group and some $3d$ theories.

\section{Superconformal index}

Superconformal index counts BPS states
protected by one supercharge which cannot be combined to form
long multiplets. The $\mathcal{N }=1$ superconformal algebra
of space-time symmetry group $SU(2,2|1)$ is generated by $J_i,
\overline{J}_i$ (Lorentz rotations), $P_\mu$ (translations),
$K_\mu$, (special conformal transformations),
$H$ (dilatations) and $R$ ($U(1)_R$-rotations). There are also four supercharges
$Q_{\alpha},\overline{Q}_{\dot\alpha}$ and their superconformal
partners $S_{\alpha},\overline{S}_{\dot\alpha}$.
Distinguishing a pair of supercharges, say,
$Q=\overline{Q}_{1 }$ and $Q^{\dag}=-{\overline S}_{1}$, one has
\begin{equation}
\{Q,Q^{\dag}\}= 2{\mathcal H},\quad Q^2=\left(Q^\dag\right)^2=0,\qquad
\mathcal{H}=H-2\overline{J}_3-3R/2.
\label{susy}\end{equation}
The SCI is defined now by the gauge invariant trace
\begin{eqnarray}  && \makebox[-2em]{}
I(p,q,\y) =  \text{Tr} \Big( (-1)^{\mathcal F}
p^{\mathcal{R}/2+J_3}q^{\mathcal{R}/2-J_3}
 \prod_{k}y_k^{F_k}\,  e^{-\beta {\mathcal H}}\Big),
\quad \mathcal{R}= R+2\overline{J}_3,
\label{Ind}\end{eqnarray}
where ${\mathcal F}$ is the fermion number operator. Parameters
$p$ and $q$ are fugacities
for the operators $\mathcal{R}/2\pm J_3$
commuting with $Q$ and $Q^{\dag}$. $F_k$ are the maximal torus
generators of the flavor group $F$ with the corresponding
fugacities $y_k$. Since relation (\ref{susy}) is preserved by the operators used in
(\ref{Ind}) only  zero modes of the operator $\mathcal{H}$ contribute to the trace.

An explicit computation of SCIs for ${\mathcal N}=1$ theories results in the prescription
\cite{Romelsberger1,Romelsberger2,Dolan:2008qi} according to which one first
composes the single particle states index
\begin{eqnarray}\nonumber &&
\text{ind}(p,q,\z,\y) = \frac{2pq - p - q}{(1-p)(1-q)}\chi_{adj, G}(\z)
\\ && \makebox[2em]{}
+ \sum_j \frac{(pq)^{R_j/2}\chi_{R_F,j}(\y)\chi_{R_G,j}(\z) -
(pq)^{1-R_j/2}\chi_{{\bar R}_F,j}(\y)\chi_{{\bar R}_G,j}(\z)}{(1-p)(1-q)}.
\label{index}\end{eqnarray}
The contribution of gauge superfields lying in the adjoint
representation of the gauge group $G_c$ is described by the first
term in (\ref{index}). The sum over $j$ corresponds to
the contribution of chiral matter superfields $\Phi_j$ transforming as the gauge
group representations $R_{G,j}$ and flavor symmetry
group representations $R_{F,j}$. The functions $\chi_{adj}(\z)$,
$\chi_{R_F,j}(\y)$ and $\chi_{R_G,j}(\z)$ are the corresponding
characters and $R_j$ are the field $R$-charges.
The variables $z_1,\ldots, z_{\text{rank}\, G_c}$ are the maximal torus
fugacities of $G_c$.
To obtain the full SCI the function $\text{ind}(p,q,\z,\y)$ is
inserted into the ``plethystic" exponential
which is averaged over the gauge group. This yields the following matrix integral
\begin{equation}\label{Ind_fin}
I(p,q,\y) =  \int_{G_c} d \mu(\z)\, \exp \bigg ( \sum_{n=1}^{\infty}
\frac 1n \text{ind}\big(p^n ,q^n, \z^n , \y^ n\big ) \bigg ),
\end{equation}
where $d \mu(\z)$ is the $G_c$-invariant measure. This formula has no rigorous
mathematical justification and the region of its applicability is not
completely determined. In \cite{Romelsberger1,Romelsberger2} it was derived basically from
the free field theory without taking into account possible complicated
dynamical effects. Therefore formally it can be applied even to the
anomalous theories.

Let us consider an example of the $s$-confining theory from
\cite{Seiberg}. Namely, take a $4d$ $\mathcal{N }=1$
SYM theory with $G_c=SU({N})$
gauge group and $SU(N_f)_l \times SU(N_f)_r \times U(1)_B$
flavor symmetry group and $N_f={N}+1$. The original (electric) theory has ${N}+1$ left
and ${N}+1$ right quarks $Q$ and $\widetilde{Q}$ lying in the
fundamental and antifundamental representations of $SU({N})$. They
have $+1$ and $-1$ baryonic charges and the $R$-charge
$R=1/({N}+1)$. The field content of the described theory is
summarized in Table \ref{t1}. The general Seiberg duality
\cite{Seiberg:d} is supposed to live in the conformal window
$3N/2 < N_f < 3N$, and we see that the duality we consider lies
outside of it.
\begin{table}
\caption{Matter content of the electric theory}
\begin{center}\label{t1}
\begin{tabular}{|c|c|c|c|c|c|}
\hline
& $SU({N})$ & $SU({N}+1)_l$ & $SU({N}+1)_r$ & $U(1)_B$ & $U(1)_R$ \\
\hline
$Q$ & $f$ & $f$ & 1 & 1 & $\frac{1}{{N}+1}$ \\
$\widetilde{Q}$ & $\overline{f}$ & 1 & $\overline{f}$
                               & $-1$ & $\frac{1}{{N}+1}$ \\
$V$ & $adj$ & 1 & 1 & 0 & 1 \\
\hline
\end{tabular}
\end{center}
\end{table}

SCI of this (``electric") theory is given by the following EHI \cite{Dolan:2008qi}:
\begin{eqnarray}\label{IE-seiberg}
&& I_E =   \kappa_{{N}} \int_{\mathbb{T}^{{N}-1}}
 \frac{\prod_{i=1}^{{N}+1} \prod_{j=1}^{{N}}
    \Gamma(s_i z_j,t^{-1}_i z^{-1}_j;p,q)}
{\prod_{1 \leq i < j \leq {N}} \Gamma(z_i z^{-1}_j,z_i^{-1}
z_j;p,q)} \prod_{j=1}^{{N}-1} \frac{d z_j}{2 \pi \textup{i} z_j},
\end{eqnarray}
where $\mathbb{T}$ denotes the unit circle with positive orientation,
$\prod_{j=1}^{{N}} z_j =1$, $|s_i|, |t_i^{-1}|<1$, and the balancing condition reads
$ST^{-1} = pq$ with
$S =\prod_{i=1}^{{N}+1}s_i,$ $T=\prod_{i=1}^{{N}+1}t_i.$
Here we introduced the parameters $s_i$ and $t_i$ as
\begin{equation}
s_i=(pq)^{R/2}vx_i, \qquad t_i=(pq)^{-R/2}vy_i,
\label{ini_var}\end{equation}
where $v$, $x_i$ and $y_i$ are fugacities for $U(1)_B$,
$SU({N}+1)_l$ and $SU({N}+1)_r$ groups, respectively, with the constraints
$\prod_{i=1}^{{N}+1}x_i=\prod_{i=1}^{{N}+1}y_i=1$, and
$$
\kappa_{{N}} = \frac{(p;p)_{\infty}^{{N}-1} (q;q)_{\infty}^{{N}-1}}{{N}!},
\qquad
(a;q)_\infty=\prod_{k=0}^\infty(1-aq^k).
$$
We use also conventions
$$
\Gamma(a,b;p,q):=\Gamma(a;p,q)\Gamma(b;p,q),\quad
\Gamma(az^{\pm1};p,q):=\Gamma(az;p,q)\Gamma(az^{-1};p,q),
$$
where
\beq \label{ellg}
\Gamma(z;p,q)= \prod_{i,j=0}^\infty
\frac{1-z^{-1}p^{i+1}q^{j+1}}{1-zp^iq^j}, \quad |p|, |q|<1,
\eeq
is the (standard) elliptic gamma function.

According to \cite{Seiberg} the dual (``magnetic") theory  is described
by colorless mesons and baryons, i.e. the dual theory has no
gauge group, but it has the same
flavor symmetry. Its description is given in terms of baryons $B$ and $\widetilde{B}$
with $U(1)_B$-charges ${N}$ and $-{N}$ and the $R$-charges ${N}/({N}+1)$.
There are also mesons of $R$-charge $2/({N}+1)$ lying in the
fundamental representation of $SU({N}+1)_l$ and antifundamental
representation of $SU({N}+1)_r$ ($M_i^j=Q_i\widetilde{Q}^j,
i,j=1,\ldots,{N}+1$). We collect all fields data in Table \ref{t2}.
\begin{table}
\caption{Matter content of the magnetic theory}
\begin{center}\label{t2}
\begin{tabular}{|c|c|c|c|c|}
\hline
& $SU({N}+1)_l$ & $SU({N}+1)_r$ & $U(1)_B$ & $U(1)_R$ \\
\hline
$M$ & $f$ & $\overline{f}$ & 0 & $\frac{2}{{N}+1}$ \\
$B$ & $\overline{f}$ & 1 & ${N}$ & $\frac{{N}}{{N}+1}$ \\
$\widetilde{B}$ & 1 & $f$ & $-{N}$ & $\frac{{N}}{{N}+1}$ \\
\hline
\end{tabular}
\end{center}
\end{table}

The SCI of the magnetic theory is
\begin{eqnarray}\label{IM-seiberg}
&&    I_M = \prod_{1 \leq i,j \leq {N}+1} \Gamma(s_i t^{-1}_j;p,q)
\prod_{i=1}^{{N}+1} \Gamma(S s_i^{-1}, T^{-1} t_i;p,q).
\end{eqnarray}

As discovered in \cite{Dolan:2008qi},
the equality of SCIs $I_E=I_M$ coincides with the mathematical
identity initially established for ${N}=2$ in \cite{S1} as
the evaluation formula for an elliptic beta integral
and conjectured for general ${N}$ in \cite{S2}
and proven completely in \cite{Rains,spi:short}.

Following Seiberg \cite{Seiberg} one can integrate out one flavor and
come to supersymmetric quantum chromodynamics  theory with $N_f={N}$ when the classical moduli space
is modified at the quantum level leading to the chiral symmetry breaking.
From the SCI point of view the condition of integrating out a flavor
is expressed by the following constraint on fugacities
$$
s_{{N}+1} t_{{N}+1}^{-1} = pq.
$$
Substituting this restriction into (\ref{IE-seiberg}) and
using the reflection property
$$
\Gamma(z,\frac{pq}{z};p,q) = 1,
$$
we see that the gamma functions involving parameters $s_{{N}+1}$ and $t_{{N}+1}$
disappear, while the expression for the dual theory (\ref{IM-seiberg})
seems to vanish, since $\Gamma(pq;p,q)=0$. However, this is true only
for generic values of parameters $s_i$ and $t_i, i=1,\ldots,{N},$ and
a more accurate analysis of the corresponding SCIs should be carried
out for special values of the fugacities.
Namely, if $s_it_j^{-1}\to 1$ for some $i$ and $j$, then
$\Gamma(s_it_j^{-1};p,q)$ diverges and we have two competing regimes.
Resolution of the emerging uncertainty can lead to a non-zero answer.
The naive prescription for building SCIs \eqref{index} and \eqref{Ind_fin}
does not apply in such cases.

\section{Chiral symmetry breaking for $G_c=SU(2)$} \label{ChSymBrSU(2)}

We start our analysis of SCIs for the case of $4d$ $\N=1$ SYM
theory with $SU(2)$
gauge group and four quark fields ($N_f=2$) considered in \cite{Seiberg}.
As we will see,
SCIs vanish for generic values of fugacities and in some special cases
they have delta function type singularities.

Let us take four parameters $s_j\in\T$ subject to the balancing
constraint $\prod_{j=1}^4 s_j=1$. In the parametrization
$s_j=e^{2\pi\textup{i}\phi_j},\, 0\leq \phi_j< 1$, one has $\sum_{j=1}^4\phi_j=0
\; (\textup{mod}\; 1)$.
Denote as $\T_d$ an infinitesimal deformation of the unit circle with positive
orientation such that the points $s_j$ lie inside $\T_d$ and the points
$s_j^{-1}$ are outside $\T_d$. Particular values of $s_j$ when such a contour does not
exist represent a special interest and they will be treated through a limiting procedure.
For this set of parameters we define the integral
\beq \label{c1_e}
I_E = \frac{(p;p)_\infty (q;q)_\infty}{2}
\int_{\T_d} \frac{\prod_{j=1}^4 \Gamma(s_j z^{\pm 1};p,q)}
{\Gamma(z^{\pm 2};p,q)} \frac{dz}{2 \pi \textup{i}z}.
\eeq
Our aim is to show that for arbitrary values of parameters $s_j$
(excluding the cases when $s_j=s_k$ for $j\neq k$)
one can evaluate this integral and come to
the equality $I_E=I_M$ with
\beq \label{c1_m}
I_M = \frac{1}{(p;p)_\infty (q;q)_\infty}
\sum_{j=2}^4\Gamma(s_1s_k, s_1s_l, s_js_k, s_js_l;p,q)\,
\delta(\phi_1 +\phi_j),
\eeq
where the triple $(j,k,l)$ is a cyclic permutation of $(2,3,4)$
and $\delta (\phi)$ is the periodic Dirac delta function with period 1,
$\delta(\phi+1)=\delta(\phi)$.
There are many equivalent forms of $I_M$, e.g.
$$
I_M =
\Gamma(s_1^{\pm1}s_2^{\pm1};p,q)\frac{\delta(\phi_1 +\phi_3)
+\delta(\phi_1 +\phi_4)}{(p;p)_\infty (q;q)_\infty}
+\Gamma(s_2^{\pm1}s_3^{\pm1};p,q)\frac{\delta(\phi_1 +\phi_2)}
{(p;p)_\infty (q;q)_\infty}.
$$
It can be checked that $I_M$ is symmetric in parameters $s_j$
due to the balancing condition $\prod_{j=1}^4 s_j = 1$,
although this is not apparent. The equality $I_E=I_M$
can be obtained by taking accurate limits of parameters in the elliptic
 beta integral which will be described below. We observe from the
above relation that for generic values of $s_j, j=1,\ldots,4,$ expressions
 $I_E$ and $I_M$ vanish and only for the cases when $s_js_k=1$ for
some $j\neq k$ one has a non-trivial result.

Univariate functions determined by Cauchy type integrals are holomorphic except of the
values of the argument lying on the integration contour. Corresponding simple pole
singularities lead to branch cuts and the function may have different values for
the argument approaching the integration contour from one or another side
as described by the Sokhotsky-Plemelj formulas \cite{mus}. Different singular kernels
can lead to more complicated  integral singularities.
Positions of zeros and poles of elliptic gamma functions indicate that
the integral \eqref{c1_e} defines a holomorphic function of parameters for $|s_j|<1$.
As follows from general considerations of \cite{Rains} the product of this
integral with the function
$\prod_{1\leq i<j\leq 4}\prod_{k,l=0}^\infty(1-s_is_jp^kq^l)$ is a
holomorphic function for all values of $s_j$. This means that the integral
may be singular only for domains of $s_j$ determined by zero locus of
the indicated multiplier. Our aim is to determine the nature of such singularities
for $s_is_j=1$, $i\neq j$, and show that they are described by Dirac delta functions.

To make formulas (\ref{c1_e}) and (\ref{c1_m}) a little more transparent
and lucid, let us assume that we deal with the singular manifold
for delta function $\delta(\phi_1+\phi_3)$. This means
that $s_1s_3=1$ which also implies that $s_2s_4=1$ because
of the balancing condition. As a result, one has in the
numerator of the integrand of $I_E$ the following expression
\beq \label{num1}
\Gamma(s_1^{\pm1}z^{\pm1}, s_2^{\pm1}z^{\pm1};p,q),
\eeq
and the coefficient depending only on elliptic gamma functions in $I_M$
has the form
\beq \label{num2}
\Gamma(s_1^{\pm1}s_2^{\pm1};p,q).
\eeq
The structure of these products of elliptic gamma functions
suggests the physical meaning of the above identity $I_E=I_M$
as the equality  of SCIs for the taken theory with the
chiral symmetry breaking and its dual.

So, the $4d$ $\N=1$ SYM theory with $SU(2)$ gauge group and two
(left and right) flavors has a naive $SU(4)$ flavor symmetry group.
The gauge invariant combinations of chiral fields are
$$
V^{ij} = Q^i Q^j,
$$
where $Q^i, i=1,\ldots,4,$ are chiral superfields in the fundamental
representation of the gauge group $SU(2)$. They are restricted at the
classical level by the following relation
\beq
\epsilon_{i_1 i_2 i_3 i_4} V^{i_1 i_2} V^{i_3 i_4} \ = \ 0.
\eeq
At the quantum level this restriction is deformed and becomes
the nonzero pfaffian constraint
\beq
Pf\, V  =  \Lambda^4,
\eeq
where $\Lambda$ is some characteristic energy scale.
This scale breaks the conformal symmetry and, so,
the term ``superconformal index" is misleading
in this case, i.e. it can be called like that only by its origin
being a supersymmetric index for a nonconformal
theory on the $S^3\times S^1$ manifold \cite{FS}.

Classical $SU(4)$ flavor symmetry group gets broken to $SP(4)$ \cite{Seiberg}
due to the modified quantum mechanical constraint. So, the true flavor
symmetry group is $SP(4)$. Apart from the chiral symmetry breaking occurring
for $N_f=2$ flavors it happens that the original electric theory can be
described at low energies in terms of the free fields
determined by gauge invariant operators. One has here the
confinement with the chiral symmetry breaking which differs from
the $s$-confinement \cite{Csaki:1996zb,Csaki:1996zb2} studied
from the SCI technique point of view in \cite{Dolan:2008qi,SV1}.
The matter content of
both electric and magnetic theories is described in Table \ref{t3},
where we put the matter content of both electric and magnetic theories
in two subtables one atop of another. The upper one gives the matter content
of the electric theory while the lower one reproduces the
confining magnetic theory (which does not have the gauge fields).

\begin{table}\label{sp4}
\caption{Matter content of two descriptions of SYM theory with
$G_c=SU(2)$ and 4 quarks}
\begin{center}\label{t3}
\begin{tabular}{|c|c|c|c|}
  \hline
   & $SU(2)$ & $SP(4)$ & $U(1)_R$ \\
\hline
  $Q$ & $f$ & $f$ & 0 \\
  $V$ & $adj$ & 1 & 1 \\
 \hline \hline
  $q$ & & $T_A$ & 0 \\ \hline
\end{tabular}
\end{center}
\end{table}

Let us comment on what we have found so far.
At the quantum level the original theory with $SU(4)$ flavor
symmetry is not complete at the arbitrary point in moduli space
due to the quantum mechanical constraint. In this case SCI
is equal to zero which is described by the relation $I_E=I_M=0$
for generic values of the fugacities $s_i$.
The points of moduli space where the  chiral symmetry breaking
occurs bring the proper quantum gauge theory with its confining
phase described by the dual theory of free chiral superfields.
They are related to the special fugacity values for which
SCIs diverge instead of vanishing.
Based on this property we conclude that the equality $I_E=I_M$
describes the chiral symmetry breaking phenomenon for supersymmetric
quantum chromodynamics  theory with
$N_f=2$ flavors. It should be stressed that although the formal
expression for SCI of the electric theory (\ref{c1_e}) is built
using the general prescription for constructing SCIs, the integration
measure should be chosen in a rather careful way (one cannot use
the contour $\T$ in  (\ref{c1_e})). Moreover, for the magnetic
theory (\ref{c1_m}) even the formal expression of SCI \textit{cannot}
be derived using this prescription due
to the appearance of the delta functions. A naive application of the
general prescription in this case would produce
infinity for the magnetic SCI, which is easily seen from
the character of the absolutely antisymmetric tensor
representation $T_A$ of $SP(4)$:
\beq \label{char}
\chi_{T_A, SP(4)} \ = \ s_1s_2+s_1s_2^{-1}+s_1^{-1}s_2
+s_1^{-1}s_2^{-1} + 1.
\eeq
The constant $1$ entering this expression formally produces the
diverging factor
$\Gamma(1;p,q)$ from the plethystic exponential (which means that
the sum in the exponential diverges). Very formally
one can interpret $\Gamma(1;p,q)$ as the value of one of
 the delta functions in (\ref{c1_m}) when its argument vanishes.
Consider again the case when $\phi_1+\phi_3=0$ implying
$\phi_2+\phi_4=0$, which lead to $s_1s_3=s_2s_4=1$. Then
in the expression (\ref{num1}) one easily recognizes
the contribution from the character of fundamental
representation of $SP(4)$
$$
\chi_{f, SP(4)} \ = \ s_1+s_1^{-1}+s_2+s_2^{-1},
$$
and in the expression (\ref{num2}) one sees only a part of the
character of the $T_A$-repre\-sen\-ta\-tion (\ref{char}).

In \cite{SV3,SV3b} we already faced the fact that the prescription
of computing SCIs in the form given in \cite{Romelsberger2,Dolan:2008qi}
requires modifications in some particular cases. Here we also see that
 it does not cover theories with the quantum deformed moduli space.
It would be nice to understand how this difficulty emerges
from the localization procedure used for computing SCIs,
which we do not discuss here.

\section{$SU({N}),\; {N}>2$, gauge group case}

Consider now the general ${N}$ case. According to \cite{Seiberg}
there are two different ways of getting the confinement with chiral symmetry
breaking for ${N}>2$. At the classical  level one has the following constraint
\beq \label{const_cl}
\det M - B \widetilde{B} \ = \ 0,
\eeq
where mesons $M$ and baryons $B,\widetilde{B}$ are defined as
\beqa
M^{i}_{\widetilde{i}} &=& Q^i \widetilde{Q}_{\widetilde{i}},
\quad i, \widetilde{i}=1,\ldots,{N},
\nonumber \\
B &=& \frac{1}{{N}!} \epsilon_{i_1 \ldots i_{{N}}} Q^{i_1} \ldots Q^{i_{{N}}},
\nonumber \\ \widetilde{B} &=& \frac{1}{{N}!} \epsilon_{\widetilde{i}_1
\ldots \widetilde{i}_{{N}}} \widetilde{Q}^{\widetilde{i}_1} \ldots
\widetilde{Q}^{\widetilde{i}_{{N}}}.
\eeqa

In \cite{Seiberg} it was shown that the classical constraint (\ref{const_cl})
is deformed quantum mechanically due to the one instanton effects to
\beq \label{const_quant}
\det M - B \widetilde{B} \ = \ \Lambda^{2{N}},
\eeq
where $\Lambda$ is some scale. Again, one has broken conformal
symmetry and our index requires an appropriate interpretation
in the context of non-conformal theories.

\subsection{Breaking to the diagonal subgroup:
$SU({N})_l \times SU({N})_r \rightarrow SU({N})_d$.} \label{BreakingSU}

Condition (\ref{const_quant}) can be resolved by fixing
\beqa
&& B = \widetilde{B} = 0, \qquad
M^{i}_{\widetilde{i}} = \Lambda^2 \delta^i_{\widetilde{i}},
\quad i, \widetilde{i}=1,\ldots,{N},
\eeqa
which leads to breaking of the flavor symmetry
$SU({N})_l \times SU({N})_r$ to the diagonal subgroup $SU({N})_d$.
As a result one has the dual theories presented in Table \ref{t4}.
\begin{table}
\caption{Matter content of two descriptions of SYM theory with
$G=SU({N})$ and  $2{N}$ quarks with the symmetry
breaking $SU({N})_l \times SU({N})_r \rightarrow SU({N})_d$}
\begin{center}\label{t4}
\begin{tabular}{|c|c|c|c|c|}
  \hline
   & $SU({N})$ & $SU({N})_d$ & $U(1)_B$ & $U(1)_R$ \\
\hline
  $Q$ & $f$ & $f$ & 1 & 0 \\
  $\widetilde{Q}$ & $\overline{f}$ & $\overline{f}$ & $-1$ & 0 \\
  $V$ & $adj$ & 1 & 0 & 1 \\
 \hline \hline
  $M$ & & $adj$ & 0 & 0 \\ \hline
  $S_1$ & & 1 & ${N}$ & 0 \\ \hline
  $S_2$ & & 1 & $-{N}$ & 0 \\ \hline
\end{tabular}
\end{center}
\end{table}

The electric theory SCI has the form
\beq \label{c2_e} I_E  =
\kappa_{{N}}
\int_{\mathbb{T}_d^{{N}-1}} \frac{\prod_{i,j=1}^{{N}}
\Gamma( u e^{2\pi\textup{i} \theta_i}
z_j,u^{-1} e^{- 2\pi\textup{i} \chi_i} z_j^{-1};p,q)}{\prod_{1 \leq i < j \leq
{N}} \Gamma(z_iz_j^{-1},z_i^{-1}z_j;p,q)} \prod_{j=1}^{{N}-1}
\frac{dz_j}{2 \pi \textup{i} z_j},
\eeq
while the magnetic SCI is
\beqa   &&\makebox[-1em]{}
I_M =  \frac{\Gamma(u^{\pm {N}};p,q) }
{ (p;p)_\infty^{{N}-1}(q;q)_\infty^{{N}-1} }
\prod_{1\leq i<j\leq {N}}\Gamma(e^{\pm 2\pi\textup{i} (\theta_i-\theta_j)};p,q)
\sum_{\tilde\theta_j}
\prod_{i=1}^{{N}-1}\delta(\chi_i - \tilde{\theta}_i),
 \label{c2_m}\eeqa
where the sum goes over permutations of parameters
$(\tilde\theta_1,\ldots,\tilde\theta_{{N}})=
\mathcal{P}(\theta_1,\ldots,\theta_{{N}})$.

The equality $ I_E = I_M$ is proved in the following section.
Here we use parametrization of fugacities
in the exponential form and
$$
\prod_{i=1}^{{N}} z_i =  1, \qquad
\sum_{i=1}^{{N}} \theta_i = \sum_{i=1}^{{N}} \chi_i \ = \ 0.
$$
For $\theta_i=\chi_i$ one can easily recognize in $I_E$
contributions of the characters of respective electric theory
field representations as described in Table \ref{t4}.
As to the magnetic theory, the meson field $M$ described by
the adjoint representation and respective character
$$
\chi_{adj, SU({N})}
=\sum_{i,j=1}^{{N}}e^{2\pi\textup{i}(\theta_i-\theta_j)}-1,
$$
yields the $\theta_j$-dependent term in \eqref{c2_m} multiplied
by the diverging factor $\Gamma(1;p,q)^{{N}-1}$, which formally
plays the role of the product of delta functions. The contribution
to $I_M$ of the scalar fields $S_1$ and $S_2$ is described
by the terms $\Gamma(u^{\pm {N}};p,q)$ using the standard prescription.
So, we see that the original recipe of building SCIs for
$SU({N})$ supersymmetric field theories with chiral symmetry breaking
requires
appropriate modification on both electric (namely, by correct choice of
the integration contour $\T_d$) and magnetic (by correct description of
singularities in the distributional sense) sides.

We would like to note that for $N=4$, i.e. for $SU(4)$ gauge theory
there are several dual theories as described in \cite{SV2}. This means
that a similar chiral symmetry breaking should
take place in three other interacting $SU(4)$-gauge field theories
with $N_f=4$. We have checked that the expression for SCIs
\eqref{c2_m} is invariant with respect to transformations of
elliptic hypergeometric integrals indicated in \cite{SV2}, i.e.
the latter theories have the same index. It would be interesting to
investigate moduli spaces and other physical properties of these
multiple dual theories to see how the chiral symmetry breaking arises
in them. The general physical properties of these and other more general
dualities lying outside the conformal windows described in
\cite{SV2} are not investigated appropriately yet.
As a correction, we mention that vanishing of SCIs
stated in \cite{SV2} is wrong in general -- there are singularities
describing interesting physics and all corresponding SCIs
should be reconsidered from this point of view.

\subsection{Breaking of $U(1)_B$.} \label{BreakingBarGr}
The constraint (\ref{const_quant}) can be resolved also by fixing
\beqa
B = -\widetilde{B} = \Lambda^{{N}}, \qquad
M^{i}_{\widetilde{i}} = 0, \quad i,\widetilde i=1,\ldots, {N},
\eeqa
which leads to breaking of the $U(1)_B$-symmetry. As a result in the infrared
fixed point one gets the dual field description as described in Table \ref{t5}.
\begin{table}
\caption{Matter content of two descriptions of SYM theory with $G_c=SU({N})$
and $2{N}$ quarks with broken $U(1)_B$}
\begin{center}\label{t5}
\begin{tabular}{|c|c|c|c|c|}
  \hline
   & $SU({N})$ & $SU({N})_l$ & $SU({N})_r$ & $U(1)_R$ \\
\hline
  $Q$ & $f$ & $f$ & 1 & 0 \\
  $\widetilde{Q}$ & $\overline{f}$ & 1 & $\overline{f}$ & 0 \\
  $V$ & $adj$ & 1 & 1 & 1 \\
 \hline \hline
  $M$ & & $f$ & $\overline{f}$ & 0 \\ \hline
  $S$ & & 1 & 1 & 0 \\ \hline
\end{tabular}
\end{center}
\end{table}

The electric SCI has the same form
\beq \label{c3_e} I_E \ = \ \kappa_{{N}}
\int_{\mathbb{T}_d^{{N}-1}} \frac{\prod_{i,j=1}^{{N}} \Gamma(x_i u
z_j,y_i^{-1} u^{-1} z_j^{-1};p,q)}{\prod_{1 \leq i < j \leq {N}}
\Gamma(z_iz_j^{-1},z_i^{-1}z_j;p,q)} \prod_{j=1}^{{N}-1} \frac{dz_j}{2
\pi \textup{i} z_j},
\eeq
with $\prod_{i=1}^{{N}} z_i  = 1$,
$\prod_{i=1}^{{N}} x_i = \prod_{i=1}^{{N}} y_i = 1$,
while the magnetic index has a different form
\beq \label{c3_m}
I_M = \prod_{i,j=1}^{{N}} \Gamma(x_iy_j^{-1};p,q) \frac{ \delta({N}\varphi)}{(p;p)_\infty
(q;q)_\infty},
\eeq
where $\varphi$ is a real variable appearing from
the exponential parametrization of the $U(1)_B$-fugacity,
$u=e^{2\pi\textup{i} \varphi}$.
The equality $I_E= I_M$ in this case is also proven in the next section.
Again, for discrete values of the fugacity $u$, $u^{{N}}=1$,
the electric index uses a nontrivial modification of the
integration contour $\T_d$ with clear contribution from
characters of the fundamental ($Q$-field) and
antifundamental ($\widetilde Q$-field) representations.
In the magnetic case the situation is trickier. The tensor
meson field $M$ character yields the $x_i, y_i$-dependent
part of the expression \eqref{c3_m} whereas the delta function
$\delta({N}\varphi)$ is modelled by the character of the
scalar field $S$ which has zero charges with respect to all
groups and so yields $\Gamma(t;p,q)$ at $t=1$ which plays
the role of the delta function, i.e. a modification of
the original recipe of building SCIs is needed in this case as well.

Again, for $N=4$ similar chiral symmetry breaking should take
place in three other $SU(4)$-gauge group dual theories
with $N_f=4$ described in \cite{SV2}. We have checked that
the expression for magnetic SCI \eqref{c3_m} is invariant with
respect to transformations of EHIs given in \cite{SV2}, i.e.
these theories have the same index.

\section{Proofs}

\subsection{$N_f={N}=2$ case} Let us prove relations for SCIs
presented in the previous section. Consider first the chiral
symmetry breaking in $\mathcal{N }=1$ SYM theory with $SU(2)$ gauge
group and $N_f=2$ flavors. Properties of SCIs in this particular
case were discussed in \cite{spi:cont,S5}, but here we would like
to give an independent consideration.

Take the $s$-confining theory with the same gauge group $SU(2)$
and $N_f=3$ flavors studied in \cite{Seiberg}. Note that
all $s$-confining theories were thought
to be classified in \cite{Csaki:1996zb,Csaki:1996zb2}, however other examples of such
theories were discovered in \cite{SV1} using the SCI technique.
Since for $G_c=SU(2)$ the fundamental and antifundamental representations
are equivalent, the flavor group extends to $SU(6)$ and quark fields unify
to its fundamental representation.
Denoting $t_{1,2,3}^{-1}=s_{4,5,6}$ in \eqref{IE-seiberg}
we come to the electric SCI
\beq \label{el2}
I_E  = \frac{(p;p)_\infty(q;q)_\infty}{2}\int_{\mathbb{T}}
\frac{\prod_{j=1}^6\Gamma(s_jz^{\pm1};p,q)}
{\Gamma(z^{\pm2};p,q)}\frac{dz}{2\pi\textup{i} z}
\eeq
where $|s_j|<1$ and $\prod_{j=1}^6s_j=pq$.
This integral is known as the elliptic beta integral
and its evaluation \cite{S1} yields the magnetic SCI (cf. with \eqref{IM-seiberg})
\beq \label{mag2}
I_M  = \prod_{1\leq j<k\leq 6}\Gamma(s_js_k;p,q).
\eeq

It is possible to reduce the equality $I_E=I_M$ to $N_f=2$
case by taking the limit
\beq \label{lim} s_5s_6  =  pq e^{\epsilon}, \quad \epsilon \rightarrow 0.
\eeq
Indeed, from the inversion relation for $\Gamma(z;p,q)$ one has
$\Gamma(s_5z,s_6z^{-1};p,q) \to 1$ and integral \eqref{el2}
simplifies to \eqref{c1_e}, where the integration contour should
be inevitably deformed due to the balancing condition $s_1s_2s_3s_4=1$
emerging for $\epsilon=0$.

Let $\epsilon$ be a small positive number. Denote
\begin{equation}\label{pars}
s_1=\alpha w, \quad s_2=\alpha w^{-1},\quad
 s_3=\beta y,\quad  s_4=\beta y^{-1}.
\end{equation}
Then
\beq\label{redm}
I_M  =
\Gamma(s_5s_6, \alpha^2, \beta^2, \alpha \beta w^{\pm1}
y^{\pm1};p,q)\prod_{m=5}^6\Gamma(\alpha s_{m} w^{\pm1}, \beta s_{m} y^{\pm1};p,q).
\eeq
Because
\beq \label{limGam}
\Gamma(s_5s_6;p,q) \stackreb{=}{\epsilon\to0} \epsilon (p;p)_\infty
(q;q)_\infty + \texttt{O}(\epsilon^2),
\eeq
the integral $I_E$ \eqref{el2} is proportional to $\epsilon$ and,
for generic values of other parameters, it vanishes for
$\epsilon =0$. However, for special values of $\alpha, \beta, w,$ and $y$
one has poles in \eqref{redm} and corresponding singularities
may alter the integral value. Let us consider the situation
for singular points $w=y$ (corresponding to $s_1s_4=1$),
$w=y^{-1}$ (corresponding to $s_1s_3=1$), and $\alpha^2=1$
(corresponding to $s_1s_2=1$).

The balancing condition $s_5 s_6 \alpha^2 \beta^2 = pq$ can be written in the form
\beq \label{limab}
\alpha\beta =e^{-\epsilon/2}.
\eeq
Actually, on the right hand side of \eqref{limab} one may have the minus
sign, but it can be removed by the change $y\to -y$ in the original
notation and therefore we stick to the positive sign in \eqref{limab}.
Keeping $\epsilon >0$ one
has $|\alpha\beta|<1$, i.e. it is possible to choose $|\alpha|, |\beta|<1$.
Suppose that $w, y\in \T$, i.e. $|w|, |y|=1$. Then all parameters
of the elliptic beta integral are of modulus less than 1 and its
evaluation \eqref{redm} holds true.

To see the nature of singularities emerging at $y=w^{\pm1}$ let us
assume that $\alpha^2\neq 1$,
multiply $I_E$ by a function $f(y)$ holomorphic near the unit circle
and integrate over the variable $y$:
\beqa \label{addint}
&&  \int_{\mathbb{T}}
f(y)I_E \frac{dy}{2 \pi \textup{i} y}
=\Gamma(s_5s_6,\alpha^2, \beta^2;p,q)
\prod_{m=5}^6\Gamma(\alpha s_{m} w^{\pm1};p,q)
\nonumber \\ && \makebox[2em]{} \times
\int_{\mathbb{T}} f(y)
\Gamma(\alpha \beta w^{\pm1} y^{\pm1};p,q)
\prod_{m=5}^6\Gamma(\beta s_{m} y^{\pm1};p,q)
\frac{dy}{2 \pi \textup{i} y}.
\eeqa

Since we may keep absolute values of $s_5$ and $s_6$
sufficiently small, so that $|\beta s_{5,6}|<1$,
there are no problems with the integration of
$s_{5,6}$-dependent elliptic gamma functions in \eqref{addint}.
The term $\Gamma(\alpha \beta w^{\pm1} y^{\pm1};p,q)$ has
the following sequences of poles and zeros
\begin{itemize}
\item {\text poles:} $\quad  y_{in} = \alpha \beta
                   w^{\pm1} p^{i} q^{j}$, \quad
$y_{out} = \frac{1}{\alpha \beta} w^{\pm1} p^{-i} q^{-j}$;
\item {\text zeros:} $\quad y= \frac{1}{\alpha \beta} w^{\pm1}
p^{i+1} q^{j+1}, \alpha \beta  w^{\pm1} p^{-i-1} q^{-j-1}$,
                 \end{itemize}
where $i,j \in \mathbb{Z}_{\geq 0}$ and the {\it in}-poles
converge to zero $y=0$ and {\it out}-poles go to infinity.
Because of the taken constraints the unit circle separates
{\it in} and {\it out} poles for $\epsilon>0$. However,
for $\epsilon\to 0$ one has $\alpha\beta\to 1$ and two
pairs of poles at $ \alpha \beta w^{\pm1}$ and
$w^{\pm1}/{\alpha \beta}$ start to pinch $\T$.  Therefore we deform
the contour of integration $\T$ to $\mathcal C$ such that only the
residues of the $ \alpha \beta w^{\pm1}$-poles are picked up
during this deformation and there are no singularities lying on
$\mathcal C$. These distinguished poles are simple for $w^2\neq 1$
(i.e., $s_1\neq s_2$),
which is assumed in the following. As a result we obtain
\beqa \label{addint_res}
&&  \int_{\mathbb{T}}
f(y)I_E \frac{dy}{2 \pi \textup{i} y}
=\Gamma(s_5s_6,\alpha^2, \beta^2;p,q)
\prod_{m=5}^6\Gamma(\alpha s_{m} w^{\pm1};p,q)
\nonumber \\ && \makebox[2em]{} \times
\Biggl(\frac{f(\alpha\beta w)}{(p;p)_\infty(q;q)_\infty}
\Gamma((\alpha\beta)^2,(\alpha\beta w)^2,w^{-2};p,q)
\prod_{m=5}^6\Gamma(\beta s_m\alpha\beta w, \frac{s_m}{\alpha w};p,q)
\nonumber \\ && \makebox[2em]{}
+\frac{f(\alpha\beta w^{-1})}{(p;p)_\infty(q;q)_\infty}
\Gamma((\alpha\beta)^2,(\alpha\beta w^{-1})^2,w^{2};p,q)
\prod_{m=5}^6\Gamma(\beta s_m\alpha\beta w^{-1}, \frac{s_mw}{\alpha};p,q)
\nonumber \\ && \makebox[2em]{}
+ \int_{\mathcal C} f(y)
\Gamma(\alpha \beta w^{\pm1} y^{\pm1};p,q)
\prod_{m=5}^6\Gamma(\beta s_{m} y^{\pm1};p,q)
\frac{dy}{2 \pi \textup{i} y}\Biggr),
\nonumber \eeqa
where we used the relation
$$
\lim_{x\to 1}(1-x)\Gamma(x;p,q)=\frac{1}{(p;p)_\infty(q;q)_\infty}.
$$
The residue factors $\Gamma((\alpha\beta)^2;p,q)$ diverge
for $\epsilon\to 0$, but $\Gamma(s_5s_6,(\alpha\beta)^2;p,q)=1$
and we obtain the finite product. However, the integral over
$\mathcal C$ remains finite and its product with
$\Gamma(s_5s_6;p,q)$ vanishes. As a result, for
$\epsilon=0$  we obtain
\beqa \label{addintfin}
&&  \int_{\mathbb{T}}
f(y)I_E \frac{dy}{2 \pi \textup{i} y}
=\frac{\Gamma(\alpha^{\pm2}, w^{\pm2};p,q)}{(p;p)_\infty(q;q)_\infty}
(f(w)+f(w^{-1}).
\eeqa

Denote $y=e^{2\pi\textup{i} \theta}, w=e^{2\pi\textup{i} \chi}$
and pass to the integration over real variable $\theta$.
Because of the arbitrariness of the function $f(y)$, we can
give to the function $I_E$ a distributional sense and write
\beqa \label{fin1}
&&  \frac{(p;p)_\infty(q;q)_\infty}{2}\int_{\mathbb{T}_d}
\frac{\Gamma(\alpha w^{\pm1}z^{\pm1},\alpha^{-1} y^{\pm1}z^{\pm1};p,q)}
{\Gamma(z^{\pm2};p,q)}\frac{dz}{2 \pi \textup{i} z}
\\ \nonumber && \makebox[2em]{}
=\frac{\Gamma(\alpha^{\pm2}, w^{\pm2};p,q)}{(p;p)_\infty(q;q)_\infty}
(\delta(\theta+\chi)+\delta(\theta-\chi)),
\eeqa
where $\delta(\theta)$ is the $1$-periodic Dirac delta function.
Note that the limit $\epsilon\to 0$ inevitably enforces the change
of the integration contour in $I_E$ from $\T$ to $\T_d$,
which is a deformation of $\T$ such that the sequences of poles
$\alpha w^{\pm1} p^iq^j$ and $\alpha^{-1}y^{\pm1}p^iq^j$
with $i,j\in\Z_{\geq 0}$ lie inside $\T_d$ and their reciprocals ---
outside of this contour. If $|\alpha|<1$ then some poles
from the second set lie outside $\T$, i.e. the contour
deformation is not infinitesimal.
For symmetric functions $f(z)=f(z^{-1})$,
the equality  \eqref{fin1} has an interpretation as
the inversion relation for an integral operator
introduced in \cite{spi:bailey}, which was demonstrated
in \cite{SW}. In turn, it was identified in \cite{DSumn} with one
of the Coxeter relations for permutation groups.

Consider now the singularities of $I_E$ at $\alpha^2=1$.
For that we multiply $I_E$ by a holomorphic function $f(\alpha)$
and integrate over $\alpha$ along the contour $\mathcal C$ which
is a deformation of $\T$ near the points $\alpha=\pm1 $
such that it passes in between the points $\alpha=1$
and $\alpha=e^{-\epsilon/2} $ on the one side and points $\alpha=-1$ and
$\alpha=-e^{-\epsilon/2} $ on the other side. Again, in the limit
$\epsilon \to 0$ two pairs of poles pinch the integration
contour and we deform $\mathcal C$ to an infinitesimal deformation
of $\T$ such that both points $\alpha=\pm 1$ lie inside it,
and pick up $\alpha=\pm1$ pole residues.
Repeating considerations similar to the previous case we obtain
\beqa \nonumber
&&  \int_{\mathcal C}
f(\alpha)I_E \frac{d\alpha}{2 \pi \textup{i} \alpha}
=\Gamma(s_5s_6,e^{-\epsilon/2}w^{\pm1}y^{\pm1};p,q)
 \int_{\mathcal C}f(\alpha)\Gamma(\alpha^2,e^{-\epsilon}\alpha^{-2};p,q)
\\ \nonumber && \makebox[2em]{} \times
\prod_{m=5}^6\Gamma(\alpha s_{m} w^{\pm1},
e^{-\epsilon/2}\alpha^{-1}s_{m} y^{\pm1} ;p,q) \frac{d\alpha}{2 \pi \textup{i} \alpha}
\\ && \makebox[2em]{}
\stackreb{=}{\epsilon\to 0}\frac{\Gamma(w^{\pm1}y^{\pm1};p,q)}
{(p;p)_\infty(q;q)_\infty}\frac{f(1)+f(-1)}{2}.
\label{res_alpha}\eeqa

Denote $\alpha=e^{2\pi \textup{i}\varphi}$ and pass to the integration
over the real variable $\varphi\in[0,1[$. Then we can write in
the distributional sense
\beqa \label{fin_alpha}
&&  \frac{(p;p)_\infty(q;q)_\infty}{2}\int_{\mathbb{T}_d}
\frac{\Gamma(\alpha w^{\pm1}z^{\pm1},\alpha^{-1} y^{\pm1}z^{\pm1};p,q)}
{\Gamma(z^{\pm2};p,q)}\frac{dz}{2 \pi \textup{i} z}
\\ \nonumber && \makebox[2em]{}
=\frac{\Gamma(w^{\pm1}y^{\pm1};p,q)}
{(p;p)_\infty(q;q)_\infty}\frac{\delta(\varphi)+\delta(\varphi-1/2)}{2}.
\eeqa
Equivalently we can write
$\delta(\varphi)+\delta(\varphi-1/2)=2\delta(2\varphi)$, because of the
periodicity of the delta function.
For these considerations to be valid we have to assume that
$y^{\pm1}w^{\pm1}\neq 1$, i.e. the previously considered
regime of parameters and the current one should not overlap,
which we assumed.

Return now from notation \eqref{pars} to the original one
$s_i=e^{2\pi \textup{i}\phi_i}$,
which means that $\phi_1=\varphi+\theta, \phi_2=\varphi-\theta,
\phi_3=-\varphi+\chi,\phi_4=-\varphi-\chi$. Then the arguments
of our delta functions are $2\varphi=\phi_1+\phi_2$,
$\theta+\chi=\phi_1+\phi_3$, and $\theta-\chi=\phi_1+\phi_4$.
Therefore summing right-hand sides of \eqref{fin1} and \eqref{fin_alpha}
we come to the expression for magnetic SCI \eqref{c1_m}
which we wanted to prove. Note that due to our constraints on the
parameters the supports of three delta functions do not overlap.

\subsection{The higher rank case,  $N_f={N}>2$}
Consider the general case $N_f={N}>2$. The situation with
breaking $SU(N_f)_l \times SU(N_f)_r \times U(1)_B \rightarrow
SU(N_f)_d \times U(1)_B$ is similar to the one described
above for $y=w^{\pm1}$.
Analysis of the $U(1)_B$-breaking, $SU(N_f)_l \times SU(N_f)_r \times U(1)_B
\rightarrow SU(N_f)_l \times SU(N_f)_r$, is
analogous to the investigation of $\alpha^2=1$ singularities above.

We start from the $s$-confining theory with $SU({N})$ gauge group
with $N_f={N}+1$ flavors. The electric theory SCI is
 \beq
I_E = \kappa_N \int_{\mathbb{T}^{{N}-1}}
\frac{\prod_{i=1}^{{N}+1} \prod_{j=1}^{{N}} \Gamma((pq)^{\frac{1}{2({N}+1)}}
x_i v z_j,(pq)^{\frac{1}{2({N}+1)} }y_i^{-1} v^{-1} z_j^{-1};p,q)}
{\prod_{1 \leq i < j \leq {N}}
\Gamma(z_iz_j^{-1}, z_i^{-1}z_j;p,q)} \prod_{j=1}^{{N}-1}
\frac{dz_j}{2 \pi \textup{i} z_j},\eeq
where $\prod_{j=1}^{{N}+1}x_j=\prod_{j=1}^{{N}+1}y_j=1$,
so that the balancing condition is satisfied automatically.
It admits exact evaluation yielding the magnetic theory SCI
\beqa
I_M =
\prod_{i=1}^{{N}+1} \Gamma((pq)^{\frac{{N}}{2({N}+1)}}x_i^{-1} v^{{N}},
(pq)^{\frac{{N}}{2({N}+1)}}y_i  v^{-{N}};p,q) \prod_{i,j=1}^{{N}+1}
\Gamma((pq)^{\frac{1}{{N}+1}}x_iy_j^{-1};p,q).
\eeqa

As in the previous considerations we would like to integrate out one
flavor by taking the limit $(pq)^{\frac{1}{{N}+1}}x_{{N}+1}y_{{N}+1}^{-1}
=pqe^{\epsilon}$, $\epsilon \rightarrow 0$.
Introduce new variables $a_i, b_i,$ and $u$:
$$
x_i=\frac{a_i}{x_{{N}+1}^{1/{N}}}, \quad
y_i=\frac{b_i}{y_{{N}+1}^{1/{N}}}, \quad i=1,\ldots, {N},
\quad  v=(pq)^{-\frac{1}{2({N}+1)}}x_{{N}+1}^{1/{N}}u,
$$
which will play the role of fugacities for $N_f={N}$ reduced theory,
$\prod_{i=1}^{{N}}a_i=\prod_{i=1}^{{N}}b_i=1$.
Then the indices take the form
 \beqa &&
I_E = \kappa_N \int_{\mathbb{T}^{{N}-1}}
\frac{\prod_{i,j=1}^{{N}} \Gamma(a_i u z_j,
e^{-\epsilon/{N}}b_i^{-1}u^{-1} z_j^{-1};p,q)}
{\prod_{1 \leq i < j \leq {N}}
\Gamma(z_iz_j^{-1}, z_i^{-1}z_j;p,q)}
\nonumber \\ && \times
 \prod_{j=1}^{{N}} \Gamma(x_{{N}+1}^{\frac{{N}+1}{{N}}}u z_j,
y_{{N}+1}^{-\frac{{N}+1}{{N}}}u^{-1} z_j^{-1}e^{-\epsilon/{N}};p,q)
 \prod_{j=1}^{{N}-1}\frac{dz_j}{2 \pi \textup{i} z_j}
\eeqa
and
\beqa \nonumber &&
I_M = \Gamma(pqe^{\epsilon},u^{{N}},e^{-\epsilon}u^{-{N}};p,q)
\prod_{i,j=1}^{{N}} \Gamma(e^{-\epsilon/{N}}a_ib_j^{-1};p,q)
\\    \nonumber && \makebox[2em]{} \times
\prod_{i=1}^{{N}} \Gamma(x_{{N}+1}^{\frac{{N}+1}{{N}}}a_i^{-1} u^{{N}},
e^{-\epsilon}y_{{N}+1}^{-\frac{{N}+1}{{N}}}b_i u^{-{N}};p,q)
\\ && \makebox[2em]{} \times
\prod_{i=1}^{{N}} \Gamma(pqe^{\epsilon} y_{{N}+1}^{\frac{{N}+1}{{N}}}
b_i^{-1},pqe^{\epsilon} x_{{N}+1}^{-\frac{{N}+1}{{N}}}a_i;p,q).
\label{IMep}\eeqa

In the limit $\epsilon\to 0$ this expression vanishes for generic
values of the parameters. The singular manifold of fugacities
requiring special consideration is determined by the poles
of elliptic gamma functions in \eqref{IMep}. The fugacity
$x_{{N}+1}$ (or $y_{{N}+1}$) is an arbitrary variable and we
can keep corresponding poles away from $\T$.
Therefore for fugacities $a_i, b_i,$ and $u$ near
the unit circle the only singular points of interest are
$a_i=b_j$ and $u^{{N}}=1$. In order to
see the structure of singularities in the first case, we multiply
$I_E$ (or $I_M$) by a holomorphic function $f(b_1,\ldots, b_{{N}})$
and integrate over the variables $b_1,\ldots,b_{{N}-1}\in\T$.
The multipliers $\Gamma(e^{-\epsilon/{N}}a_ib_j^{-1};p,q)$
have the poles
$$
{in:} \quad b_j=e^{-\epsilon/{N}}a_i,
\quad i=1,\ldots,{N},\; j=1,\ldots, {N}-1,
$$
lying inside $\T$, and
$$
{out:} \quad b_{{N}}^{-1}=b_1\ldots b_{{N}-1}=e^{\epsilon/{{N}}}a_i^{-1},
\quad i=1,\ldots,{N},
$$
lying outside $\T$ for any particular $b_k$. Since $a_i\in\T$, for
$\epsilon\to 0$ all {\it in} poles approach
$\T$ from inside. Positions of the {\it out} poles
depend on the order of integration in $b_i$ and their
values. Suppose we integrate first over $b_1$, then $b_2$, etc.
Then for the pole $b_1=e^{\epsilon/{N}}/a_jb_2\ldots b_{N-1}$
there exist such values of $b_2,\ldots,b_{{N}-1}\in\T$ that
$b_1=e^{\epsilon/{N}}a_k$ for $k=1,\ldots, {N}$ and for
$\epsilon\to0$ we have pinching of the integration contour
near the points $b_1=a_k$. To escape such a pinching we shrink
the integration contours a little to pick up the residues of the
$b_i=e^{-\epsilon/ {N}}a_i$ poles lying inside $\T$
(like in the ${N}=2$ case).
After taking sequentially ${N}-1$
``residues of residues" in integration variables,
say at the point $b_i=a_i$,
on the last step  we obtain the term
$$
\Gamma(e^{-\epsilon/ {N}}a_{{N}}b_{{N}}^{-1};p,q)
=\Gamma(e^{-\epsilon};p,q),
$$
which diverges and, being multiplied by $\Gamma(pqe^{\epsilon};p,q)$,
yields the finite answer.
Evidently, we can take residues in arbitrary
possible order $b_i=e^{-\epsilon/ {N}}a_j, j=1,\ldots, {N},$
each of which yields different final result. Only the
highest order residues survive in the limit $\epsilon\to0$ since
all lower order residues vanish
due to the multiplier $\Gamma(pqe^{\epsilon};p,q)$.
As a result we obtain
\beqa \label{SUN1} &&
\int_{\T^{{N}-1}}
f(b_1,\ldots,b_{{N}})
I_M\prod_{j=1}^{{N}-1}\frac{db_j}{2\pi\textup{i} b_j}
\stackreb{=}{\epsilon\to 0}
\frac{\Gamma(u^{\pm {N}};p,q)}{(p;p)_\infty^{{N}-1}(q;q)_\infty^{{N}-1}}
\\ \nonumber && \makebox[2em]{}  \times
\prod_{1\leq i<j\leq {N}}\Gamma(a_ia_j^{-1},a_i^{-1}a_j;p,q)
\sum_{\tilde a_j} f(\tilde a_1,\ldots, \tilde a_{{N}}),
\eeqa
where summation goes over all permutations of parameters
appearing from different orders of taking residues,
$(\tilde a_1,\ldots, \tilde a_{{N}})=\mathcal{P}(a_1,\ldots,a_{{N}})$.

To tackle the singularities at $u^N=1$ we multiply $I_M$ by
a holomorphic function $f(u)$ and integrate over $u$ along
the contour which is an infinitesimal deformation of $\T$
passing in between the points $u=e^{2\pi\textup{i}k /{N}}$
and $u=e^{-\epsilon/{N}}e^{2\pi\textup{i}k /{N}},$ $k=0,\ldots, {N}-1$.
Then we deform the integration contour and pick up the residues
at $u=e^{2\pi\textup{i}k /{N}}$. For $\omega^{{N}}=1$ one has
$$
\lim_{u\to \omega} (1-\omega/u)\Gamma(u^{{N}};p,q)
=-\frac{1}{{N}(p;p)_\infty(q;q)_\infty}.
$$
The contours for computing residues are oriented
clockwise, which results in the extra minus sign and yields
\beq
\int_{\T_d}f(u)I_M\frac{du}{2\pi\textup{i} u}
=\prod_{1\leq i,j\leq {N}}\Gamma(a_ib_j^{-1};p,q)
\frac{1}{{N}(p;p)_\infty(q;q)_\infty}
\sum_{k=0}^{{N}-1}f(e^{2\pi\textup{i}k /{N}}).
\label{uN}\eeq

Introducing the angular variables $a_i=e^{2\pi \textup{i} \theta_i}$,
$b_i=e^{2\pi \textup{i} \chi_i}$ and $u=e^{2\pi \textup{i} \varphi}$,
we can write in the distributional sense
 \beqa\nonumber  &&
\kappa_N \int_{\mathbb{T}_d^{{N}-1}}
\frac{\prod_{i,j=1}^{{N}} \Gamma(e^{2\pi \textup{i}(\theta_i+\varphi)} z_j,
e^{-2\pi\textup{i}(\chi_i+\varphi)} z_j^{-1};p,q)}
{\prod_{1 \leq i < j \leq {N}} \Gamma(z_iz_j^{-1}, z_i^{-1}z_j;p,q)}
\prod_{j=1}^{{N}-1}\frac{dz_j}{2 \pi \textup{i} z_j}
\\ \nonumber &&  \makebox[2em]{}
=\frac{\Gamma(u^{\pm {N}};p,q)}{(p;p)_\infty^{{N}-1}(q;q)_\infty^{{N}-1}}
 \prod_{1\leq i<j\leq {N}}\Gamma(a_ia_j^{-1},a_i^{-1}a_j;p,q)
\sum_{\tilde \theta_j}\prod_{k=1}^{{N}-1}\delta(\chi_k-\tilde\theta_k)
\\ &&  \makebox[2em]{}
+\prod_{1\leq i,j\leq {N}}\Gamma(a_ib_j^{-1};p,q)
\frac{\delta({N}\varphi)}{(p;p)_\infty(q;q)_\infty},
\label{sunindfin}\eeqa
where the sum $\sum_{\tilde \theta_j}$ goes over all ${N}!$
permutations of the variables $(\theta_1,\ldots,\theta_{{N}})$
and $\delta({N}\varphi)=(1/{N})\sum_{k=0}^{{N}-1}\delta(\varphi-k/{N})$.
This is a general formula describing simultaneously both
cases of chiral symmetry breaking.

Interestingly, for $N=4$ the expression \eqref{sunindfin}
has an extended symmetry generated by the reflection of
fugacities and multiplication by some elliptic gamma functions
described in \cite{SV2} in association with three
more dual theories with nontrivial $SU(4)$-gauge group
interaction.

\subsection{The case $N_f<{N}$}
Take the electric theory with $G_c=SU(2)$ and a single flavor $N_f=1$.
Corresponding  SCI has the form
\beq
\frac{(p;p)_\infty(q;q)_\infty}{2}\int_{\mathcal C}
\frac{\Gamma((pq)^{-\frac 12}
e^{\pm \textup{i} \theta} z^{\pm1};p,q)}{\Gamma(z^{\pm2};p,q)}
\frac{dz}{2 \pi \textup{i} z},
\eeq
where the integration contour $\mathcal C$ separates the poles
converging to zero from their reciprocals.
It can be formally obtained from the $N_f=2$ index by setting
$\alpha=\sqrt{pq}$. For generic values of $\theta$ this integral
vanishes, as a consequence of the elliptic beta integral evaluation.
However, it is not completely clear for which values of $\theta$
there are singularities allowing one to obtain a non-zero answer
in the distributional sense. It is not legitimate to simply
substitute $\alpha=\sqrt{pq}$ into \eqref{fin1} since that relation
was obtained under the condition $|\beta s_5|, |\beta s_6|<1$.
For $\alpha \to \sqrt{pq}$ one has $\beta^2\to (pq)^{-1}$,
so that $s_5s_6\beta^2\to 1$ and there emerge additional pinchings
of the $y$-variable integration contour.

Consider SCI for the pure $SU(2)$ SYM theory, i.e. the $N_f=0$ case.
This theory has $R$-symmetry anomaly and the corresponding
SCI is described not by an EHI, but by a theta hypergeometric
integral \cite{S2} (i.e. no balancing condition is present):
\beq
I_{pure, SU(2)} = \frac{(p;p)_\infty(q;q)_\infty}{2}
\int_{\mathbb{T}} \frac{1}{\Gamma(z^{\pm2};p,q)}
 \frac{dz}{2 \pi \textup{i} z}.
\label{Nf=0}\eeq
As was mentioned already, this formula emerges from the formal free field
considerations. Actually, absence of the $R$-symmetry makes it questionable
how to define the superconformal index on the $S^3\times S^1$ manifold.
Therefore it is not completely clear what kind of data are described
by the expression \eqref{Nf=0}.

Nevertheless, the integral \eqref{Nf=0} can be evaluated explicitly. To compute it, we use the
inversion formula for  elliptic gamma functions
$$
\frac{1}{\Gamma(z^{\pm2};p,q)} = \theta(z^2;p) \theta(z^{-2};q),
$$
where the theta function is defined as
$$
\theta(z;p) = (z;p)_\infty (pz^
{-1};p)_\infty = \frac{1}{(p;p)_\infty}
\sum_{k \in \mathbb{Z}} (-1)^k p^{k(k-1)/2} z^k.
$$
Applying the latter series expansion for theta functions
twice we get
\beqa
&& I_{pure\, SU(2)} =  \frac 12 \sum_{k,l \in \mathbb{Z}} (-1)^{k+l}
p^{k(k-1)/2} q^{l(l-1)/2} \int_{\mathbb{T}} z^{2(k-l)}
\frac{dz}{2 \pi \textup{i} z} \nonumber \\ && \makebox[4.5em]{}
= \frac 12 \sum_{k \in \mathbb{Z}} ( pq)^{k(k-1)/2}
= \frac{1}{2}(pq;pq)_\infty \theta(-1;pq).
\label{Nf0}\eeqa
Using the plethistic exponential we can also write
\beq
I_{pure\, SU(2)} =
 (pq;pq)_\infty(-pq;pq)_\infty^2
=\exp \Big(-\sum_{n=1}^\infty\frac{(pq)^n+2(-pq)^n}{n(1-(pq)^n)}\Big).
\label{Nf0plet}\eeq
The physical meaning of this relation is not completely clear.
Perhaps, the right-hand side expression in \eqref{Nf0plet}
hints on the formation of the gaugino condensate \cite{gaugino,gaugino2}.

\section{Chiral symmetry breaking for  $G_c=SP(2{N})$}

Consider chiral symmetry breaking
in a $\N=1$ SYM theory with the gauge group  $SP(2{N})$. Let us start
from the $s$-confining theory
with $G_c=SP(2{N})$ and $2{N}+4$ quarks studied in
\cite{Intriligator:1995ne} with the identification of the number
of flavors as $N_f=N+2$. Corresponding (electric)
SCI is \cite{Dolan:2008qi,SV1}
\begin{eqnarray}\label{sp1_1}
&& \makebox[-2em]{}
I_E=\frac{(p;p)^{{N}}_{\infty} (q;q)^{{N}}_{\infty}}{2^{{N}}{N}!}
\int_{\mathbb{T}^{{N}}} \prod_{1 \leq i < j \leq {N}}
\frac{1}{\Gamma(z_i^{\pm 1} z_j^{\pm 1};p,q)}
\nonumber \\ && \makebox[2em]{} \times \prod_{j=1}^{{N}}
\frac{\prod_{m=1}^{2{N}+4}
\Gamma(t_mz_j^{\pm 1};p,q)}{\Gamma(z_j^{\pm 2};p,q)}
\frac{d z_j}{2 \pi \textup{i} z_j},
\eeqa
where $|t_m|<1$ and
the balancing condition reads $\prod_{m=1}^{2{N}+4} t_m = pq.$
The dual (magnetic) theory is described by colorless mesons
forming the $T_A$-representation of $SU(2{N}+4)$ group
with the index
\beqa\label{sp1_2} &&
I_M= \prod_{1 \leq m < s \leq 2{N}+4} \Gamma(t_mt_s;p,q).
\end{eqnarray}
The equality $I_E=I_M$ was suggested in \cite{DS} and
proved in \cite{Rains,spi:short}.
As in the previous $G_c=SU(N)$ case we integrate out
two quark fields by restricting chemical potentials as, say,
$t_{2{N}+3}t_{2N+c+4}=pq$. As a result, dependence on the parameters
$t_{2{N}+3}$ and $t_{2{N}+4}$ disappears from $I_E$ which yields formally
the index of the theory with $2{N}+2$ chiral fields.
For generic values of other fugacities, $I_M$ is equal to zero,
but there are delta function singularities
for a singular submanifold of fugacities.
For the taken $SP(2N)$-gauge group the conformal window where the
general Seiberg duality is supposed to be valid has the form
$3(N + 1)/2 < N_f < 3(N + 1)$, our duality corresponds to $N_f=N+1$
and lies outside this window.

A theory with $SP(2{N})$ gauge group and quantum modified moduli space
was described in \cite{Intriligator:1995ne}.
The matter content for corresponding electric and magnetic theories
is presented in Table \ref{t6}. The mesonic fields are composed
as $M_{ij} = Q_iQ_j$, where the $SP(2N)$ symplectic trace is assumed
making the mesons gauge invariant. The quantum moduli space of vacua
satisfies the constraint
$$
Pf\, M = \Lambda^{2(N+1)},
$$
with some energy scale $\Lambda$ which breaks the conformal symmetry
with appropriate consequences for interpreting our SCIs.

\begin{table}\label{t6}
\caption{A $4d$ theory with $G_c=SP(2{N})$
and $2{N}+2$ quarks exhibiting the chiral symmetry breaking}
\begin{center}
\begin{tabular}{|c|c|c|c|}
  \hline
   & $SP(2{N})$ & $SP(2({N}+1))$ & $U(1)_R$ \\
\hline
  $Q$ & $f$ & $f$ & 0 \\
  $V$ & $adj$ & 1 & 1 \\
\hline \hline
  $M$ &  & $T_A$ & 0 \\ \hline
\end{tabular}
\end{center}
\end{table}

Naively the electric theory has the $SU(2({N}+1))$ flavor group with $2{N}+1$
independent fugacities. Corresponding fundamental representation
character has the form
\beq
\chi_{f,SU(2{N}+2)} (\underline{x}) =
\sum_{i=1}^{2{N}+2} x_i,
\label{charSU2N2}\eeq
where $x_i$ are fugacities for maximal torus generators of
$SU(2{N}+2)$ restricted by the constraint $\prod_{i=1}^{2{N}+2} x_i = 1.$
The chiral symmetry breaking reduces this naive flavor group to
$SP(2{N})$. Therefore it is necessary to describe how the character
\eqref{charSU2N2} reduces to the fundamental representation
character of $SP(2{N})$
\beq
\chi_{f,SP(2{N}+2)} (\underline{y}) =
\sum_{i=1}^{{N}+1} (y_i+y_i^{-1}),
\label{charSP2N2}\eeq
where $y_1,\ldots, y_{{N}+1}$ are maximal torus fugacities
without constraints. Evidently, this can be done if one identifies
half of $x_i$ variables with $y_j$ and forces the rest
of $x_i$-variables to coincide with $y_j^{-1}$
(which resolves automatically the balancing condition).
This observation hints that one should realize the constraints
$x_ix_j=1,\, i\neq j,$ for all possible splittings of $x_i$-variables
into pairs.

In order to find the structure of $I_E$ in the case of chiral symmetry
breaking, we set $t_{2{N}+3}t_{2N+c+4}=pqe^{\epsilon}$
in \eqref{sp1_1} and \eqref{sp1_2} and consider the limit $\epsilon\to 0$.
Then, expression
\eqref{sp1_2} contains the multiplier $\Gamma(pqe^{\epsilon};p,q)$
tending to zero which can be overpowered only by the poles
of other elliptic gamma functions. Because now
$\prod_{j=1}^{2{N}+2}t_j=e^{-\epsilon}$, in the limit $\epsilon\to 0$
we can identify $t_j=x_j$. Originally, the equality $I_E=I_M$
was obtained for $|t_i|<1$ for all $i$, however it can be meromorphically
continued to arbitrary values of the parameters.
To test the singularities we
multiply $I_M$ by an arbitrary holomorphic function
$f(t_1,\ldots,t_{{N}+1})$ weighted by a specific product of elliptic
gamma functions and integrate over $t_1,\ldots, t_{{N}}\in\T$:
$$
\int_{\T^{{N}}}\rho(\underline{t})f(t_1,\ldots,t_{{N}+1})\, I_E
\prod_{k=1}^{{N}}\frac{dt_k}{2\pi\textup{i} t_k},\quad
\rho(\underline{t})=\frac{1}{\prod_{1\leq i< j\leq {N}+1}\Gamma(t_it_j;p,q)},
$$
where we assume that the balancing condition is
resolved in favor of the variable $t_{{N}+1}$:
$$
t_{{N}+1}=\frac{e^{-\epsilon}}{\prod_{k=1}^{{N}}t_k
\prod_{l=1}^{{N}+1}t_{l+{N}+1}}.
$$
Multiplication of $I_E$ by $\rho(\underline{t})$
removes a number of singularities which are associated
with the zero locus of $\rho(\underline{t})$. However,
the latter singularities can be restored later on by the
permutational symmetry in variables $t_i$. Replacing
$I_E$ by $I_M$ we come to the expression
\beqa\nonumber && \makebox[-2em]{}
\Gamma(pqe^{\epsilon};p,q)\prod_{{N}+2\leq i< j\leq 2{N}+2}\Gamma(t_it_j;p,q)
\int_{\T^{{N}}}f(t_1,\ldots,t_{{N}+1})
\\ \nonumber && \makebox[2em]{} \times
\prod_{i=1}^{{N}}\prod_{j={N}+2}^{2{N}+2}\Gamma(t_it_j;p,q)
\prod_{j={N}+2}^{2{N}+2}\Gamma(\frac{t_je^{-\epsilon}}
{\prod_{k=1}^{{N}}t_k\prod_{l=1}^{{N}+1}t_{l+{N}+1}};p,q)
\\ && \makebox[2em]{} \times
\prod_{j=1}^{2{N}+2}\Gamma(t_jt_{2{N}+3},\frac{pqe^{\epsilon}t_j}
{t_{2{N}+3}} ;p,q)
\prod_{k=1}^{{N}}\frac{dt_k}{2\pi\textup{i} t_k}.
\label{sp}\eeqa
Consider singularities of the integrand near the integration
contours. For $\epsilon>0$ we can take $|t_j|=e^{-\epsilon/({N}+2)}<1,\,
i={N}+1,\ldots, 2{N}+2$,
so that in the limit $\epsilon\to 0$ one has $t_i\to\T$ for
$i=1,\ldots, 2{N}+2$. Let us take the absolute values of $t_{2{N}+3}$
and $t_{2{N}+4}$ sufficiently small, so that
the poles of the elliptic gamma functions
on the last line in \eqref{sp} do not approach $\T$
and stay harmless. Then the relevant poles are
$$
out: \quad t_i=t_j^{-1}, \quad i=1,\ldots, {N},\quad
j={N}+2,\ldots, 2{N}+2,
$$
lying outside $\T$ and
$$
in: \quad \prod_{i=1}^{{N}}t_i=\frac{e^{-\epsilon}t_j}
{\prod_{l=1}^{{N}+1}t_{l+{N}+1}},\quad j={N}+2,\ldots,2{N}+2,
$$
lying inside $\T$. Consider first the integral in $t_1$.
There always exist such values of $t_2,\ldots, t_{{N}}$
that the {\em in} poles approach $\T$ from inside at the points
$t_1\to t_j^{-1},\, j={N}+2,\ldots, 2{N}+2,$
 and there emerge pinchings of $\T$ by {\em in} and {\em out} poles.
These poles are simple provided $t_j\neq t_k,\, j\neq k$, which we assume.
To deal with that we inflate a little all integration contours $\T$
and pick up the resides of all {\em out} poles. These residues
have singularities of a similar structure and one can continue
taking these ``residues of residues" in $t_2, t_3,$ etc
until the last integration variable $t_{{N}}$.
Considering the sequence of residues at $t_i=t_{{N}+1+i}^{-1}$,
on the last step one obtains the diverging multiplier
$\Gamma(t_{{N}+1}t_{2{N}+2};p,q)=\Gamma(e^{-\epsilon};p,q)$
which cancels the vanishing factor $\Gamma(pqe^{\epsilon})$.
Similar situation holds for taking pole residues in any other possible
order. In the limit $\epsilon\to 0$ only these
highest order residues survive, since if one misses at least
one residue in the intermediate step, no divergency is taking
place and the corresponding term vanishes.

As a result, we obtain
\beqa\nonumber &&
\int_{\T^{{N}}}\frac{f(t_1,\ldots,t_{{N}+1})}
{\prod_{1\leq i< j\leq {N}+1}\Gamma(t_it_j;p,q)}\, I_E
\prod_{k=1}^{{N}}\frac{dt_k}{2\pi\textup{i} t_k}
\\  \nonumber && \makebox[2em]{}
\stackreb{=}{\epsilon\to 0}
\frac{\prod_{{N}+2\leq i< j\leq 2{N}+2}\Gamma(t_it_j,
t_i^{-1}t_j,t_it_j^{-1};p,q)}
{(p;p)_\infty^{{N}}(q;q)_\infty^{{N}}}
\sum_{\tilde t_j} f(\tilde t_{{N}+2}^{-1},\ldots, \tilde t_{2{N}+2}^{-1}),
\eeqa
where $(\tilde t_{{N}+2},\ldots, \tilde t_{2{N}+2})=\mathcal{P}
(t_{{N}+2},\ldots, t_{2{N}+2})$ is any permutation of the
parameters.

Denote now $t_j=e^{2\pi\textup{i} \phi_j}$ and use real
variables $\phi_j$ to write $I_E$ as a distribution.
The full set of singularities of $I_E$, which was
partially reduced after multiplication by $\rho(\underline{t})$,
is restored from complete $S_{2{N}+2}$-group permutational
symmetry of the index.

Because of the balancing condition  $\sum_{i=1}^{2{N}+2}\phi_i=0$
we have $2{N}+1$ independent variables $\phi_i$.
Assume as before that
$\phi_{{N}+1}$ is fixed by other parameters. Consider an
arbitrary split of the set $\Phi=(\phi_1,\ldots, \phi_{2{N}+2})$
into two $({N}+1)$-term groups
$\Phi_1=(\tilde \phi_1,\ldots, \tilde \phi_{{N}},
\tilde \phi_{{N}+1}=\phi_{{N}+1})$
and $\Phi_2=(\tilde\phi_{{N}+2},\ldots,\tilde \phi_{2{N}+2}).$
Then we pair parameters in $i$-th position, $i=1,\ldots, {N}$,
in these groups
and impose the constraints $\tilde\phi_i+\tilde\phi_{{N}+1+i}=0$.
Because of the balancing condition, the remaining pair of parameters
satisfies the constraint $\phi_{{N}+1}+\phi_{2{N}+2}=0$
automatically. Now we form a sum of products of delta functions
$$
\sum_{S_{{N}+1}(\Phi_2)}\prod_{i=1}^{{N}}\delta(\tilde\phi_i+\tilde\phi_{{N}+1+i}),
$$
where the sum goes over all possible $({N}+1)!$ permutations
of elements of the set $\Phi_2$. Evidently this sum is
also symmetric under ${N}!$ permutations of the elements
in the first set $\Phi_1$ and $2^{{N}}$ permutations of
$\tilde\phi_i$ with $\tilde\phi_{{N}+1+i}$ belonging to different sets.
Using this auxiliary building block, we can write the final
relation for our SCIs in the following form
\beqa &&
I_E = \frac{(p;p)^{{N}}_\infty
(q;q)^{{N}}_\infty}{2^{{N}} {N}!} \int_{\mathbb{T}_d^{{N}}} \prod_{1
\leq i < j \leq {N}} \frac{1}{\Gamma(z_i^{\pm1}z_j^{\pm1};p,q)}
\nonumber \\ && \makebox[2em]{} \times
\prod_{j=1}^{{N}} \frac{\prod_{i=1}^{2{N}+2}\Gamma(e^{2\pi\textup{i} \phi_i}
z_j^{\pm1};p,q)}{\Gamma(z_j^{\pm2};p,q)} \frac{dz_j}{2 \pi \textup{i} z_j}
=I_M=\frac{1}{(p;p)_\infty^{{N}} (q;q)_\infty^{{N}}}
\label{spscis}\\ && \makebox[-2em]{} \times
\sum_{(\Phi_1\bigcup\Phi_2)/S_2^{{N}} }
 \prod_{1 \leq i < j \leq {N}+1}
\Gamma(e^{2\pi \textup{i} (\pm \tilde \phi_i \pm \tilde \phi_j)};p,q)
\sum_{S_{{N}+1}(\Phi_2)}
\prod_{i=1}^{{N}}\delta(\tilde\phi_i+\tilde\phi_{{N}+1+i}),
\nonumber \eeqa
where the first sum goes over all possible splits of $\Phi$ into
$\Phi_1$ and $\Phi_2$ modulo $2^{{N}}$ permutation of the
paired parameters. In the electric SCI
the integration contour $\T_d$ is a deformation of $\T$
such that it separates sequences of the integrand poles
converging to zero from their reciprocals, i.e. $e^{2\pi\textup{i}\phi_i}$
lie inside $\T_d$ and $e^{-2\pi\textup{i}\phi_i}$ are outside $\T_d$.

It is not difficult to see that one can replace fixed $\phi_{{N}+1}$ by
any other parameter and it will give the same result, i.e. the
final answer is $S_{2{N}+2}$-group symmetric. Therefore,
one may replace both sums in \eqref{spscis} by a single sum over
all permutations of $\phi_i$, $i=1,\ldots, 2{N}+2,$
and divide it by $(2{N}+2){N}!2^{{N}}$ counting
the number of equal terms.
For ${N}=1$ relation \eqref{spscis} coincides with the equality of SCIs
considered in Sect. \ref{ChSymBrSU(2)}.

We conclude that the electric SCI is non-vanishing only
on the support of indicated products of delta functions.
For each such product one has the reduction of
the character of fundamental representations of
$SU(2{N}+2)$-group down to the corresponding character of $SP(2{N}+2)$-group,
as prescribed by the chiral symmetry breaking and naive
recipe of building SCIs. On the dual side the products of
elliptic gamma functions coincide with the $\phi_i$-dependent
part of SCIs for free meson fields forming the
$T_A$-representation of $SP(2{N}+2)$-group with the character
$$
\chi_{T_A,\, SP(2{N}+2)}=\sum_{1\leq i<j\leq {N}+1}
\sum_{\mu=\pm1, \nu=\pm1}
e^{2\pi \textup{i} (\mu \phi_i +\nu \phi_j)} +{N}.
$$
The formal prescription for building SCIs would yield
from the constant ${N}$ the diverging factor $\Gamma(1;p,q)^{{N}}$,
which in our rigorous consideration is replaced by
the product of delta functions divided by
$(p;p)_\infty^{{N}}(q;q)_\infty^{{N}}$. We see again that
for theories with chiral symmetry breaking the standard
recipe of constructing SCIs requires a careful modification.

An interesting situation arises in the rank 3 case, i.e.
for the $SP(6)$-gauge group with $8$ chiral superfields.
In this case the multiple duality phenomenon takes
place, which follows from the considerations of \cite{SV1}
for the special value of the corresponding $U(1)$-group
fugacity $t=\sqrt{pq}$. These theories lie outside of
the conformal window and their content was described
in \cite{SV2}. This means that there are three more
interacting field theories with the same gauge group and 8 quarks
showing the chiral symmetry breaking whose ``superconformal"
indices should coincide with the one for our electric/magnetic
theory. However, the expression \eqref{spscis} does not
satisfy this property -- it is not invariant under the
transformation of fugacities from $W(E_7)$-group accompanied
by multiplication of the index by certain products of
elliptic gamma functions \cite{SV1}. Under these transformations
new combinations of the delta functions emerge which were
forbidden by our constraints on the parameters, i.e. a more
careful extended analysis of the situation is needed
which we postpone to a later time.

As to such extended symmetries
for indices we mention that the considerations of $W(E_7)$
and $W(E_6)$-invariant SCIs in \cite{DG,GV} should be reducible
to one more level down to the $W(F_4)$-symmetric instance.
Namely, there should exist some combination of
fugacities after multiplication by which
the combination of delta functions in $I_M$ for $N_f=N=2$
or for more general theories of \cite{SV1} should be
invariant with respect to the Weyl group $W(F_4)$. Again a
more detailed investigation of emerging singularities may
be required and the consideration
of such a possibility lies beyond the scope of the present work.

\section{$3d$ theories with chiral symmetry breaking}

Recently there was a breakthrough in investigation of
$3d$ supersymmetric field theories due to the calculation of
partition functions (see, e.g. \cite{KWY}-\cite{3db}).
As shown in  \cite{DSV} (see also \cite{GY,I})
$4d$ superconformal indices can be reduced to $3d$
partition functions which yields a reduction
of the related $4d$ Seiberg dualities to $3d$ SYM or CS theory
dualities. To our knowledge this scheme is the most
efficient way of producing $3d$ dualities after
appropriate amendment of the superpotentials \cite{3drev}.

To realize the $4d/3d$ reduction in the simplest
$s$-confining theory one considers a special limit
of the elliptic beta integral.
First one parametrizes the variables as
$$
p = e^{2 \pi \textup{i} r \omega_1}, \quad q = e^{2 \pi \textup{i} r \omega_2},
\quad s_j=e^{2 \pi \textup{i} r \phi_j},
\quad z = e^{2 \pi \textup{i} r u}
$$
and then takes the limit $r \rightarrow 0$. To simplify the integrals one
uses the Ruijsenaars limit for elliptic gamma function
\beq
\Gamma(e^{2 \pi \textup{i} r u};e^{2 \pi \textup{i} r \omega_1},
e^{2 \pi \textup{i} r \omega_2}) \stackreb{=}{r \rightarrow 0}
e^{-\pi \textup{i}(2z-\omega_1-\omega_2)/12r\omega_1\omega_2}
\gamma^{(2)}(u;\omega_1,\omega_2),
\eeq
where
\beq
 \gamma^{(2)}(u;\omega_1,\omega_2) = e^{-\frac{\pi \textup{i}}{2}
B_{2,2}(u;\omega_1,\omega_2)} \frac{(e^{2 \pi \textup{i}(u-\omega_2)/\omega_1};
e^{-2 \pi \textup{i} \omega_2/\omega_1})_\infty}{(e^{2 \pi \textup{i}
u/\omega_2};e^{2 \pi \textup{i} \omega_1/\omega_2})_\infty}\eeq
is the hyperbolic gamma function and $B_{2,2}(u;\mathbf{\omega})$
is the second order Bernoulli polynomial,
$$
 B_{2,2}(u;\mathbf{\omega}) =
\frac{u^2}{\omega_1\omega_2} - \frac{u}{\omega_1} -
\frac{u}{\omega_2} + \frac{\omega_1}{6\omega_2} +
\frac{\omega_2}{6\omega_1} + \frac 12.
$$
The following conventions are used below
$
\gamma^{(2)}(a,b;\mathbf{\omega}) :=
\gamma^{(2)}(a;\mathbf{\omega}) \gamma^{(2)}(b;\mathbf{\omega})
$
and
$
\gamma^{(2)}(a\pm u;\mathbf{\omega}) :=
\gamma^{(2)}(a+u;\mathbf{\omega})
\gamma^{(2)}(a-u;\mathbf{\omega}).
$
The function $ \gamma^{(2)}(u;\omega_1,\omega_2)$ has poles
at $u=-n\omega_1-m\omega_2$ for $n,m\in\Z_{\geq0}$,
zeros at $u=n\omega_1+m\omega_2$ for $n,m\in\Z_{>0}$
and satisfies the inversion
relation $ \gamma^{(2)}(u,\omega_1+\omega_2-u;\omega_1,\omega_2)=1$.

Taking the limit $r\to0$ along the negative imaginary axis
and assuming that Re$(\omega_1)$, Re$(\omega_2)>0$ one gets the following
reduction of the electric SCI (up to some diverging factor,
see e.g. \cite{DSV})
\begin{equation}\label{redgen}
I_E^{red} = \int_{-\textup{i} \infty}^{\textup{i} \infty}
\frac{\prod_{k=1}^6\gamma^{(2)}(\phi_k \pm u;\omega_1,\omega_2)}
{\gamma^{(2)}(\pm 2 u;\omega_1,\omega_2)} \frac{du}{2 \textup{i}
\sqrt{\omega_1 \omega_2}},
\end{equation}
where the balancing condition has the form
$\sum_{k=1}^6\phi_k=\omega_1+\omega_2$
and the integration contour separates sequences of poles
going to infinity on the right- and left-hand sides of the
imaginary axis. The magnetic theory SCI reduces to (up to the same
diverging factor as in $I_E$)
\begin{equation}\label{redmgewn}
I_M^{red} = \prod_{1\leq j< k\leq 6}\gamma^{(2)}(\phi_j+\phi_k;
\omega_1,\omega_2).
\end{equation}
Impose now the constraint $\phi_5+\phi_6=\omega_1+\omega_2+\epsilon$
and take the limit $\epsilon\to 0$. The limiting balancing
condition takes the form $\phi_1+\ldots+\phi_4=0$,
and we can take all $\phi_i$ as purely imaginary numbers,
$\phi_m=\textup{i}g_m,\, g_m\in\mathbb{R}$.
Let us apply the scheme of consideration of singularities of
the previous section in the present setting using the relation
$$
2\pi\textup{i} \lim_{g\to 0}g\,\gamma^{(2)}(\textup{i}g;\omega_1,\omega_2)
=\sqrt{\omega_1\omega_2}
$$
for computing the residues. As a result we obtain the expressions
\begin{equation}\label{rd1}
I_E^{red} = \int_{\mathcal C}
\frac{\prod_{m=1}^4\gamma^{(2)}(\textup{i}g_m \pm u;\omega_1,\omega_2)}
{\gamma^{(2)}(\pm 2 u;\omega_1,\omega_2)} \frac{du}{2 \textup{i}
\sqrt{\omega_1 \omega_2}},
\end{equation}
where the integration contour is an infinitesimal deformation
of the imaginary axis such that the poles
$u=\textup{i}g_j+n\omega_1+m\omega_2,\, n,m\in\Z_{\geq0}$,
lie to the right of $\mathcal C$ and their reciprocals
$u\to -u$ are to the left of $\mathcal C$.
The magnetic theory yields
\beqa\nonumber
I_M^{red} = \sqrt{\omega_1\omega_2}\Big(
\gamma^{(2)}(\pm \textup{i}g_1\pm \textup{i}g_2;\omega_1,\omega_2)
(\delta(g_1 +g_3)+\delta(g_1 +g_4))
\\ \makebox[2em]{}
+\gamma^{(2)}(\pm \textup{i}g_2\pm \textup{i}g_3;\omega_1,\omega_2)
\delta(g_1 +g_2)\Big),
\label{rm1}\eeqa
where $\delta(g)$ is the standard (non-periodic) delta function.
The equality $I_E^{red}=I_M^{red}$ expresses coincidence of
partitions functions of two $\mathcal{N }=2$ $3d$ theories whose
matter content is the same as in Table \ref{sp4} with the replacement
$4d\to 3d$.  This example of chiral symmetry breaking corresponds
to the $N_f={N}=2$ case duality in the considerations of \cite{3drev}.

Denote now
$$
g_1=\mu+x,\quad g_1=\mu-x, \quad
g_3=-\mu+y,\quad g_4=-\mu-y
$$
and take the limit $\mu\to +\infty$. Using the asymptotic
properties of the hyperbolic gamma function
for $\text{Im}(\omega_1/\omega_2)>0$,
\begin{eqnarray}\nonumber
&&
\lim_{u\to \infty}e^{\frac{\pi \textup{i}}{2}B_{2,2}(u;\mathbf{\omega})}
\gamma^{(2)}(u;\omega_1,\omega_2)=1, \qquad \text{for}\ \arg \omega_1 <\arg u
<\arg \omega_2+\pi,
\\ &&
\lim_{u\to \infty}e^{-\frac{\pi \textup{i}}{2}B_{2,2}(u;\mathbf{\omega})}
\gamma^{(2)}(u;\omega_1,\omega_2)=1, \qquad \text{for}\ \arg \omega_1 -\pi<\arg u
<\arg \omega_2,
\nonumber
\end{eqnarray}
we can see that $I_E^{red}=\gamma Z_E$,
where $\gamma$ is the diverging factor
 $\gamma=\exp(-2\pi\mu(\omega_1^{-1}+\omega_2^{-1}))$, and
\begin{equation}
Z_E = e^{\pi \textup{i} (x^2-y^2)/\omega_1\omega_2} \int_{-\textup{i}
\infty}^{\textup{i} \infty} \gamma^{(2)}(\pm \textup{i}x - u,\pm  \textup{i}y + u;
\omega_1,\omega_2) \frac{du}{2 \textup{i} \sqrt{\omega_1 \omega_2}}.
\end{equation}
Similarly, $I_M^{red}=\gamma Z_M$, where
\begin{equation}
Z_M = \sqrt{\omega_1 \omega_2}\gamma^{(2)}(\pm 2\textup{i}x;\omega_1,\omega_2)
(\delta(x-y)+\delta(x+y)).
\end{equation}
There are only two delta functions since the argument of the
third one goes to infinity, $g_1+g_2=2\mu\to+\infty$, i.e.
it does not give contributions. The multiplier
$e^{\pi \textup{i} (x^2-y^2)/\omega_1\omega_2}$ can be
dropped in $Z_E$, since  $Z_E$ vanishes for $x\neq \pm y$.

The identity $Z_E=Z_M$ expresses the equality of partition functions
of two dual $3d$ $\mathcal{N }=2$ supersymmetric field theories
described in \cite{Aharony:1997bx}.
The (real) electric theory has $U(1)$ gauge group and $N_f=2$ chiral
fields with the broken $U(1)_A$ symmetry and naive $SU(2)_l\times SU(2)_r$
flavor group broken to the diagonal subgroup $SU(2)$.
The magnetic theory has no local gauge group symmetry and
consists of only confined meson fields.
The matter content of these dual theories is presented in Table \ref{t7}.
\begin{table}
\caption{A $3d$ theory with the chiral symmetry breaking}
\begin{center}\label{t7}
\begin{tabular}{|c|c|c|c|}
  \hline
   & $U(1)$ & $SU(2)$ & $U(1)_R$ \\
\hline
  $Q$ & $f$ & $f$ & 0 \\
  $\tilde{Q}$ & $\overline{f}$ & $f$ & 0 \\
  $V$ & $adj$ & 1 & 1 \\
 \hline \hline
  $q$ & & $adj$ & 0 \\ \hline
\end{tabular}
\end{center}
\end{table}

Again, the original recipe of building $3d$ partition functions \cite{KWY}
requires a modification for theories exhibiting chiral symmetry
breaking --- in the electric part the contour of integration
should be chosen appropriately and in the magnetic part
contributions of constant terms in the characters of representations
yielding $\gamma(0;\omega_1,\omega_2)$ should be replaced
by delta functions. In the above example, the magnetic theory
meson fields form the adjoint representation of $SU(2)$
flavor group with the character $x^2+x^{-2}+1$. The latter
constant ``1" formally yields in $Z_M$ the factor
$\gamma(0;\omega_1,\omega_2)$, which should be replaced
in reality by $\sqrt{\omega_1\omega_2}(\delta(x-y)+\delta(x+y))$,
where $x$ and $y$ are fugacities of the naive flavor group $SU(2)_l\times SU(2)_r$.
We would like to stress that our interpretation of vanishing
partition functions differs from the one made in
\cite{Morita:2011cs} where corresponding
partition functions were equal to zero due to
the mass parameters lying in the general position.
Evidently one can proceed in a similar manner and
consider other examples of $4d$ dual theories with
chiral symmetry breaking and reduce them to $3d$ partners
exhibiting similar phenomenon \cite{3drev}. In particular,
it is possible to consider $3d$ partners of the $4d$
theories described in \cite{SV2} and find multiple dualities
with this property.

\section{Conclusion}

To conclude, as a continuation of our previous considerations
of the relation between properties of
elliptic hypergeometric integrals and superconformal indices
\cite{SV1}-\cite{SV5}, we have described how to compute these indices
in the theories with chiral symmetry breaking.
The  original prescription \cite{Kinney}-\cite{Romelsberger2}
needs modification in this case and the theory of elliptic
hypergeometric integrals yields the required recipe.
The chiral symmetry breaking mechanism is reflected
in the appearance of delta functions in the indices of
original theories with naive chiral symmetry such that
their support yields constraints on the fugacities
describing the quantum deformed moduli spaces with
real symmetries. This mechanism survives in the $4d\to 3d$
reduction simply by the reduction of corresponding
$4d$ superconformal indices to $3d$ partition functions.

The results from the analysis of SCIs or partition functions
allow one to find easily the field content of the theories.
However, a deeper physical investigation of our results is
needed. Namely, the physical meaning of the index
in this situation should be reconsidered with an explanation
of the emergence of delta functions from the localization
procedure.

\

{\bf Acknowledgments.} This work was partially supported by the
RFBR grant no. 12-01-00242 and the Heisenberg-Landau program.
We  would like to thank organizers
of the workshop ``Aspects of conformal and superconformal field
theories" at the Cambridge University in April 2012, where
preliminary results of this paper were reported.
We are indebted also to H. Osborn and N. Seiberg for
useful comments and to the anonymous referee for constructive remarks.

\end{document}